\documentclass[a4paper,11pt]{article}
\pdfoutput=1 
\usepackage{jinstpub} 

\usepackage{amsmath}
\usepackage{subcaption}

\usepackage[symbol]{footmisc}

\newcounter{daggerfootnote}

\begin{document}
\title{Evaluation of the upgraded 3-inch Hamamatsu photomultiplier for the KM3NeT Neutrino Telescope }
\author[a]{S.~Aiello}
\author[b,be]{A.~Albert}
\author[c]{A.\,R.~Alhebsi}
\author[d]{M.~Alshamsi}
\author[e]{S. Alves Garre}
\author[g,f]{A. Ambrosone}
\author[h]{F.~Ameli}
\author[i]{M.~Andre}
\author[j]{L.~Aphecetche}
\author[k]{M. Ardid}
% ...
\affiliation[a]{INFN, Sezione di Catania, (INFN-CT) Via Santa Sofia 64, Catania, 95123 Italy}
\affiliation[b]{Universit{\'e}~de~Strasbourg,~CNRS,~IPHC~UMR~7178,~F-67000~Strasbourg,~France}
\affiliation[c]{Khalifa University of Science and Technology, Department of Physics, PO Box 127788, Abu Dhabi,   United Arab Emirates}
\affiliation[d]{Aix~Marseille~Univ,~CNRS/IN2P3,~CPPM,~Marseille,~France}
\affiliation[e]{IFIC - Instituto de F{\'\i}sica Corpuscular (CSIC - Universitat de Val{\`e}ncia), c/Catedr{\'a}tico Jos{\'e} Beltr{\'a}n, 2, 46980 Paterna, Valencia, Spain}
\affiliation[f]{INFN, Sezione di Napoli, Complesso Universitario di Monte S. Angelo, Via Cintia ed. G, Napoli, 80126 Italy}
\affiliation[g]{Universit{\`a} di Napoli ``Federico II'', Dip. Scienze Fisiche ``E. Pancini'', Complesso Universitario di Monte S. Angelo, Via Cintia ed. G, Napoli, 80126 Italy}
\affiliation[h]{INFN, Sezione di Roma, Piazzale Aldo Moro 2, Roma, 00185 Italy}
\affiliation[i]{Universitat Polit{\`e}cnica de Catalunya, Laboratori d'Aplicacions Bioac{\'u}stiques, Centre Tecnol{\`o}gic de Vilanova i la Geltr{\'u}, Avda. Rambla Exposici{\'o}, s/n, Vilanova i la Geltr{\'u}, 08800 Spain}
\affiliation[j]{Subatech, IMT Atlantique, IN2P3-CNRS, Nantes Universit{\'e}, 4 rue Alfred Kastler - La Chantrerie, Nantes, BP 20722 44307 France}
\affiliation[k]{Universitat Polit{\`e}cnica de Val{\`e}ncia, Instituto de Investigaci{\'o}n para la Gesti{\'o}n Integrada de las Zonas Costeras, C/ Paranimf, 1, Gandia, 46730 Spain}
% ...
\affiliation[be]{Universit{\'e} de Haute Alsace, rue des Fr{\`e}res Lumi{\`e}re, 68093 Mulhouse Cedex, France}
% ...

\abstract{The 3-inch Hamamatsu R14374-02 photomultiplier tube is an improved version of the R12199-02 model and its successor in the construction of the KM3NeT neutrino telescope. A total of 1000 photomultipliers were analysed to assess their dark count rate, transit time spread, and spurious pulses. A subset of 200 photomultipliers were further evaluated to determine their quantum efficiency which is an essential parameter for Monte Carlo simulations of the detector response. The measurements show that R14374-02 model has better quantum efficiency homogeneity over the photocatode and better time properties than the R12199-02.
}


\maketitle
\flushbottom

\section{Introduction}

KM3NeT is a research infrastructure that comprises two deep-sea neutrino detectors located at the bottom of the Mediterranean Sea \cite{KM3TDR}. The two detectors, ARCA and ORCA, are optimised for different studies. The detection principle relies on the collection of Cherenkov photons emitted along the path of relativistic charged particles produced in neutrino interaction in or close to the detectors. The ARCA detector, which is designed for high-energy astrophysical neutrino studies, is located off the coast of Portopalo di Capo Passero in Italy at a depth of 3500 meters below sea level. ARCA currently comprises 33 deployed detection units (DUs) \cite{PPM_DU} with a total of 18414 photomultiplier tubes (PMTs). The ORCA detector, which is dedicated to the study of atmospheric neutrino oscillations and to the determination of the neutrino mass ordering, is located off the coast of Toulon in France at a depth of 2500 meters below sea level. ORCA currently has 23 deployed DUs containing a total of 10044 PMTs. When completed, the two detectors will comprise in total 345 DUs, with a total of 192510 PMTs. PMTs are essential components of both detectors, and their performance is critical to the success of the telescopes. Not only a high sensitivity of the photosensors is required but also their timing response has to fullfill requirements that guarantee a good angular resolution when the direction of the neutrino is reconstructed. Both detectors have a modular matrix-like structure with multiple detection units. Each DU comprises 18  digital optical modules (DOMs) \cite{DOM1,DOM2} attached at a fixed distance from each other. The DOMs are mechanically supported by two \textsuperscript{TM}Dyneema ropes, while an optoelectronic cable provides power to each DOM and ensures the optical connection for data collection and control. Data are transmitted to the base module (BM) mounted on the DU anchor on the sea floor. Both the ARCA and ORCA detectors are designed with the same type of DOMs. The main difference between the two detectors is the distance  between the optical modules: 36 m for ARCA and 9 m for ORCA in the vertical direction;  90 m for ARCA and 20 m  for ORCA in the horizontal direction. Each DOM comprises 31 3-inch PMTs that are enclosed inside a 17-inch pressure-resistant glass sphere. The requirements for the PMTs in the KM3NeT detectors were previously outlined in reference \cite{oldpmts,Herold} and are summarised in Table~\ref{tabella1}. In the initial phase of the construction of the KM3NeT detectors the Hamamatsu R12199-02 PMT was used, a 3-inch curved photocatode head-on type PMT with a standard bi-alkali metal photocatode and ten dynodes. In cooperation with the KM3NeT collaboration, Hamamatsu released a new and improved 3-inch photomultiplier tube named R14374-02 which has less spurious pulses. Spurious pulses in PMTs refer to signals that are not generated by the light hitting the PMT. 
\begin{table}
\center
\begin{tabular}{|c|c|}
\hline
Photocatode diameter & >72 mm \\
\hline
Nominal Voltage for gain $3\times10^6$	& 900$-$1300 V \\
\hline
Quantum Efficiency at 470~nm &	> 18\% \\
\hline
Quantum Efficiency at 404~nm &	> 25\% \\
\hline
Peak-to-Valley ratio &	> 2.0 \\
\hline
Transit Time Spread (FWHM) &	$<$ 5~ns \\
\hline
Dark count rate (0.3 spe threshold, at 20 $^\circ$C) & 2000~cps max\\
\hline
Prepulses between  $-60$~ns and $-10$~ns &  1.5\% max\\
 \hline
Delayed pulses between 15~ns and 60~ns &  5.5\% max\\
  \hline
Late afterpulses between 100~ns and 10 $\mu$s & 15\% max\\
\hline
\end{tabular}
\caption{Main characteristics of PMTs used in the KM3NeT detectors (\textbf{spe} refers to single photoelectrons and \textbf{cps} to counts per second).}
\label{tabella1} 
\end{table}
The Hamamatsu R14374-02 PMT is by now widely used in various fields such as particle physics, nuclear physics, and medical imaging. 
In this study the measurements carried out on the R14374-02 PMT are compared to the results reported in reference \cite{oldpmts}. This will ensure the compliance with the technical specifications of the KM3NeT experiment and highlight the technical enhancements. Furthermore, this study represents a milestone in testing and validating the tabletop apparatus shown in Figure \ref{qetestsetup} which is specifically dedicated to the characterisation of the quantum efficiency (QE) of photomultiplier tubes. Quantum efficiency is an important parameter characterising the PMT photocathode, which is the first active component of the detector interacting with photons.
\section{Quantum efficiency measurements}
 \label{sect:qe}
\begin{figure}[htbp]
\center
\includegraphics[width=0.7\textwidth]{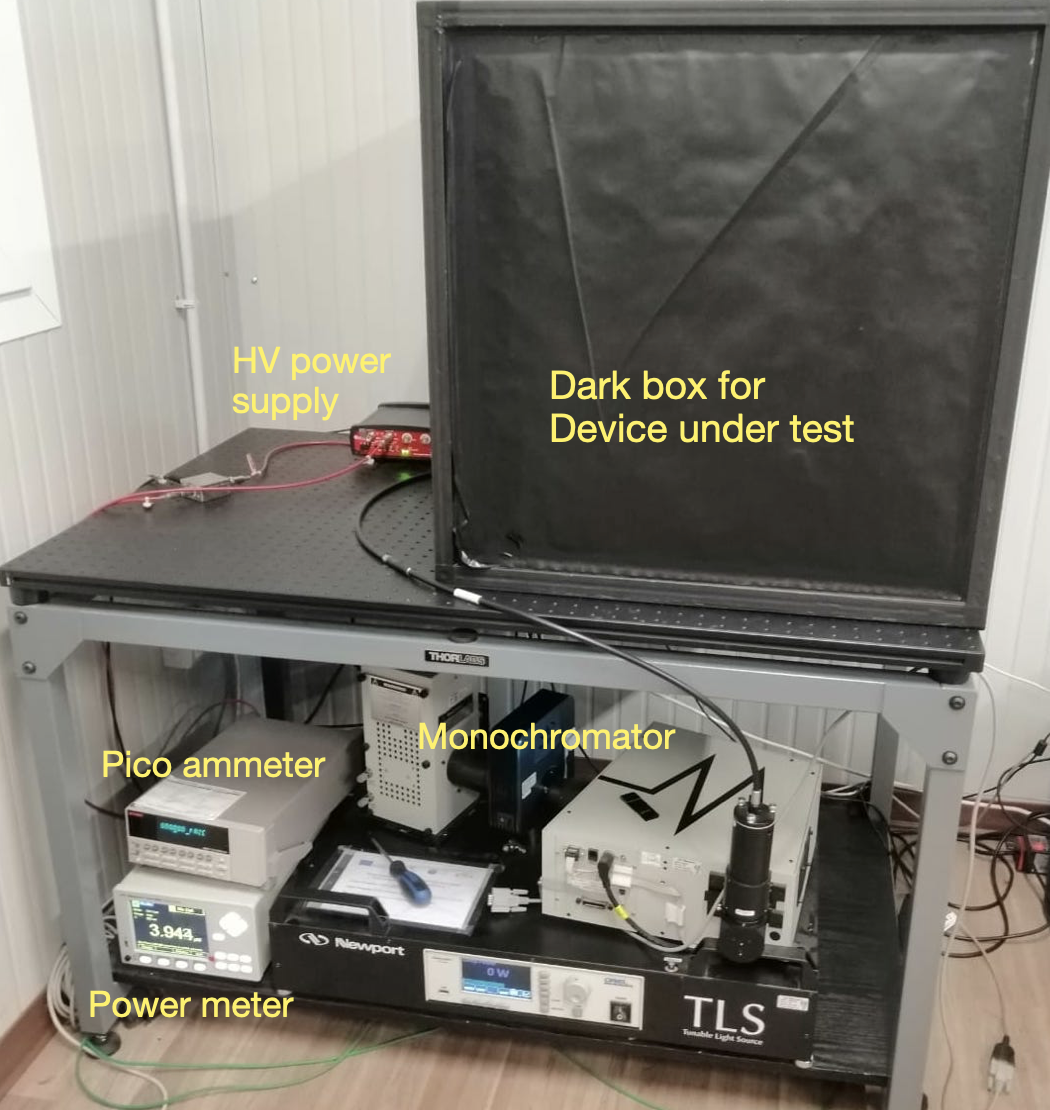}
\caption{The test setup for photocatode scan and quantum efficiency measurements at the Capacity laboratory of the Istituto Nazionale di Fisica Nucleare (INFN) located in Caserta. The dark box contains the PMT under test and the 2D moving stage. The lower part of the picture shows the Xenon lamp, the monochromator and the electronic circuits required for the data acquisition.}
\label{qetestsetup}
\end{figure}
The quantum efficiency of a photomultiplier is defined by the ratio of photoelectrons produced after the photoelectric effect takes place in the photocathode to the number of incident photons hitting the active area. It depends on the wavelength of the incoming light and should not be affected by exposure time or source fluence. 
To assess the quantum efficiency of R14374-02 photomultipliers, the apparatus shown in Figure \ref{qetestsetup} was used. The technical and methodological approach to achieve the most accurate estimation of quantum efficiency are described in detail in reference \cite{migliozzi2023scanning}.
The measurement method is based on the absolute radiometric technique. A calibrated power-meter quantifies the number of incident photons illuminating the device under test, while a calibrated picoammeter measures the resulting photocurrent. For a photomultiplier tube, the best estimation of the photocathode quantum efficiency is achieved by collecting the photoelectrons emitted via the direct photoelectric effect into the vacuum, then collected by the first dynode.
This completely automated setup enabled the acquisition of quantum efficiency measurements for a set of 200 R14374-02 photomultipliers.
\begin{figure}[htbp]
\includegraphics[width=1\textwidth]{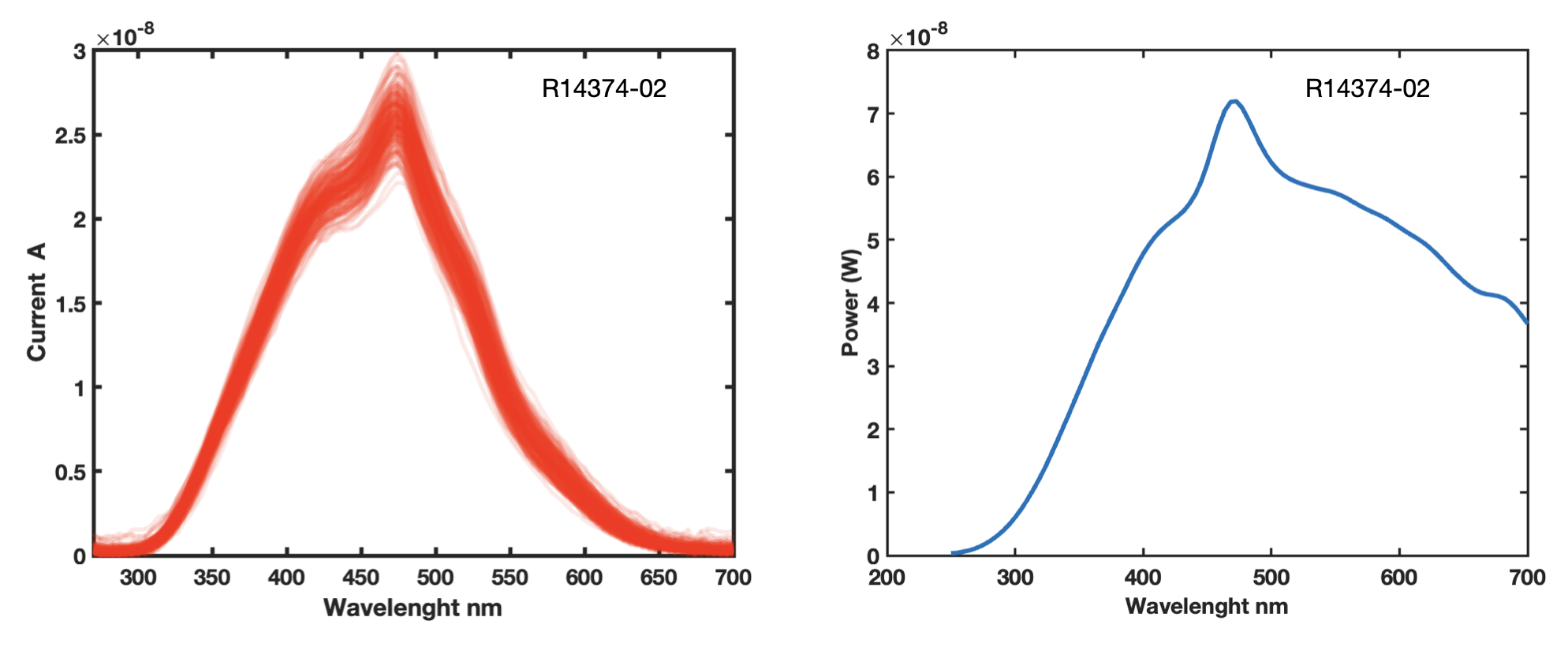}
\caption{Raw data (left panel) of photocurrent output as collected from the first dynode for 200 PMTs and the reference spectrum (right panel).\label{rawqedata}}
\end{figure}
\begin{figure}[h!]
\center
\includegraphics[width=1.0\textwidth]{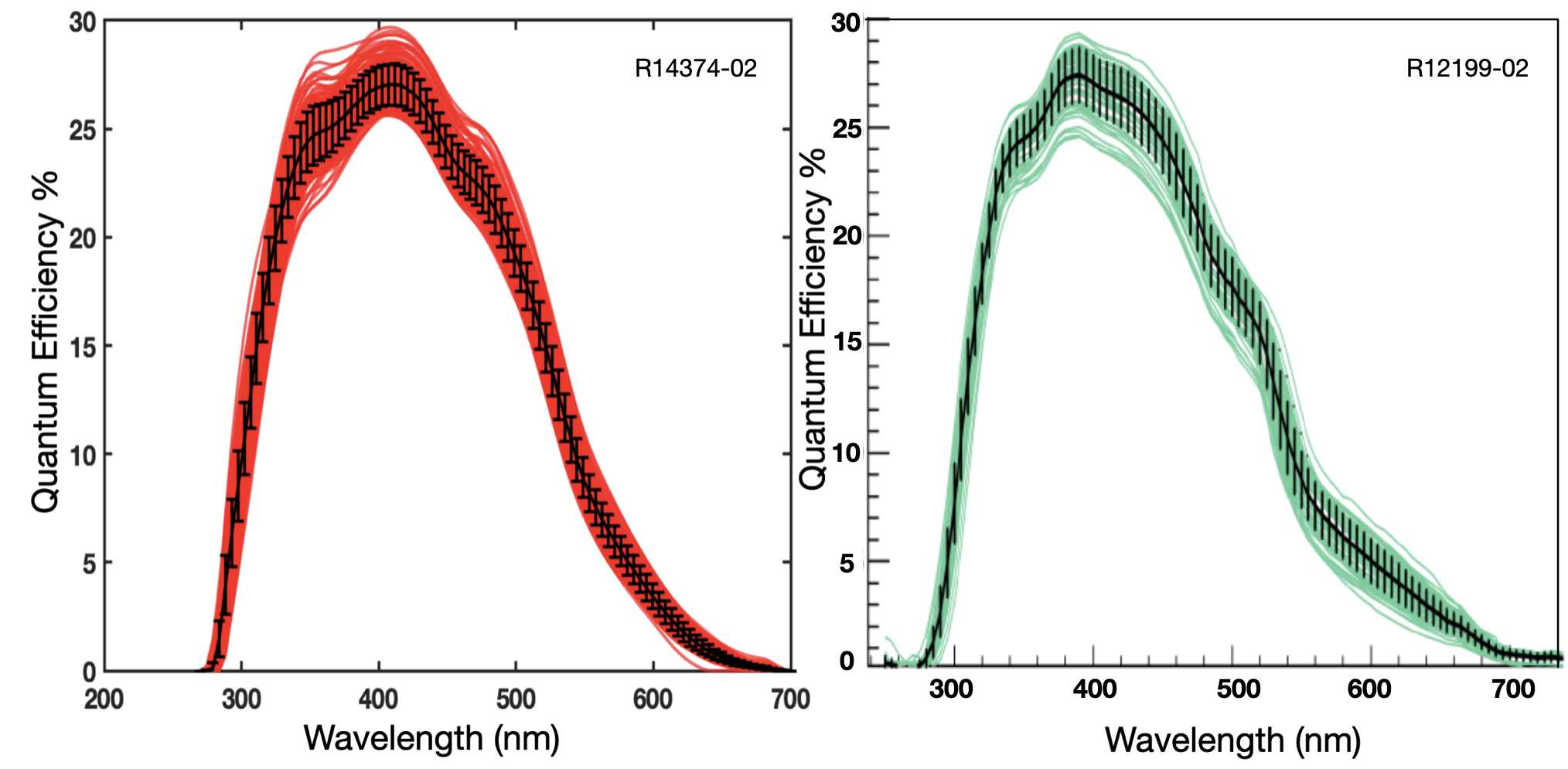}
\caption{Photocatode quantum efficiency measurements as a function of the wavelength for 200 R14374-02 (left) and 46 R12199-02 (right).}
\label{QECFR}
\end{figure}
The quantum efficiency curve is derived from the average of 200 measurements, and the error is estimated as the standard deviation of the set of measurements at a fixed wavelength. 
\subsection{Quantum efficiency results}
Individual current measurements for the 200 PMTs are shown in Figure \ref{rawqedata}. The right panel represents the reference power spectrum of the light incident on the PMT photocatode wich is used to calculate the quantum efficiency.
The mean quantum efficiency curve  shown in the left panel of Figure \ref{QECFR} was obtained averaging all measured QE values.
The errors are calculated as the standard deviation of the wavelength by wavelength measurements as described in reference \cite{oldpmts} for the Hamamatsu R12199-02 PMTs. This result is shown in the right panel for comparison. In this study not only the QE in the centre of the PMT is measured but also the radial homogeneity of the quantum efficiency. 
The two-dimensional scan of the photocatode reveals radial dishomogeneity, particularly an increase in efficiency at the outer rim due to the back reflection of incident light occurring in the internal metal structures of the PMT.  
\begin{figure}[htbp]
\center
\includegraphics[width=.55\textwidth]{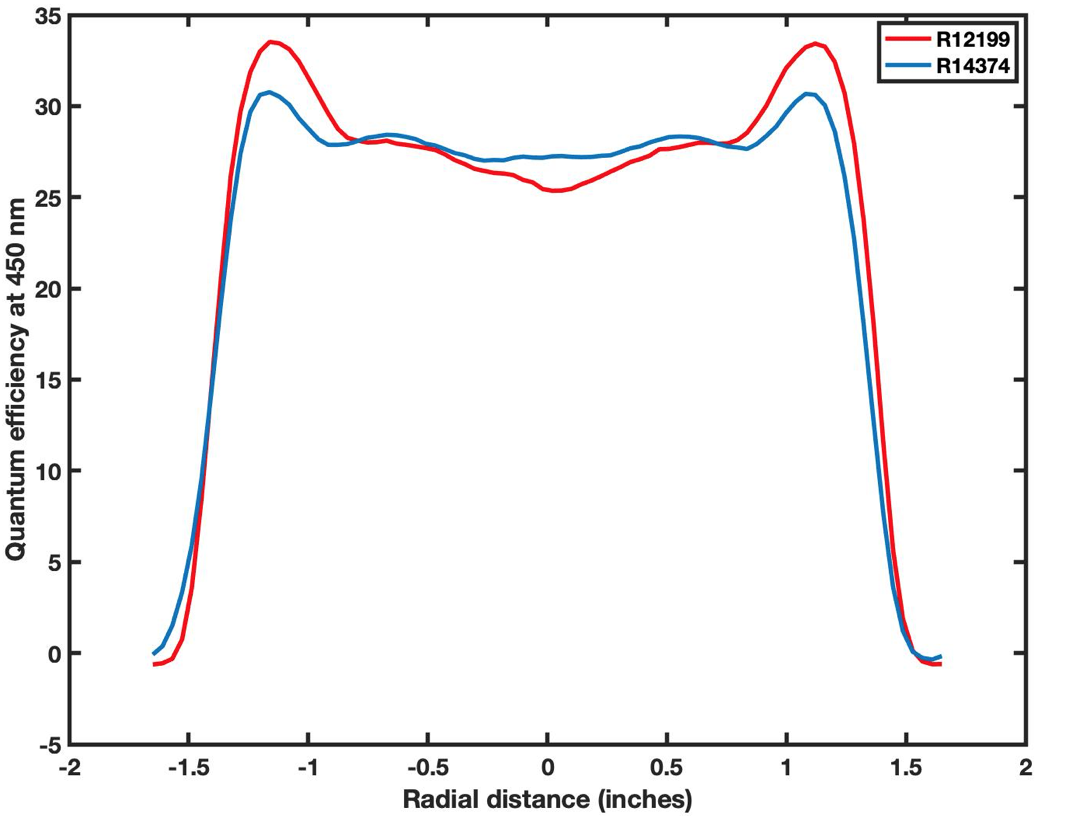}
\caption{Radial scan over the photocatode surface, showing the quantum efficiency as a function of radial distance from the center, for a wavelength of 450 nm. The blue curve shows the data collected with R14374-02 PMTs, the red curve refers to the R12199-02 PMTs measurements.}
\label{fig:qeradial}
\end{figure}
This back reflected light stimulates a secondary photoelectric emission on the photocatode. The effect is well known and described in various technical reports \cite{fotonis,hamahandbook}. The lowest value of QE is found at the centre of the active area of the PMT where back-scattered light is minimised and trapped and diffused in the dynode structure. To explore the radial distribution of the quantum efficiency the distribution was measured on a smaller set of ten R12199-02 and R14374-02 PMTs (Figure \ref{fig:qeradial}).
This investigation is performed at the wavelength of 450 nanometers.
The difference in QE at 450 nm at the photocathode centre between the two PMT types is consistent with that in Figure \ref{QECFR}; indeed at this wavelength the new model has a slighly higher QE. The results reveal a dissimilarity between the two models, with the new model displaying a reduced contrast from the edge to the centre. However, no significant improvements in QE were expected for the R14374-02 since the technology for bi-alkali photocatode deposition remains the same \cite{bialculino}.
\section{Noise, gain and time properties comparison } 
The time characteristics of a PMT are vital for photon counting applications. To assess these characteristics a set of 1000 PMTs equipped with the digital base used in KM3NeT detectors \cite{Timmer:2010zz,coating} was examined.
\begin{figure}[htbp]
\center
\includegraphics[width=0.7\textwidth]{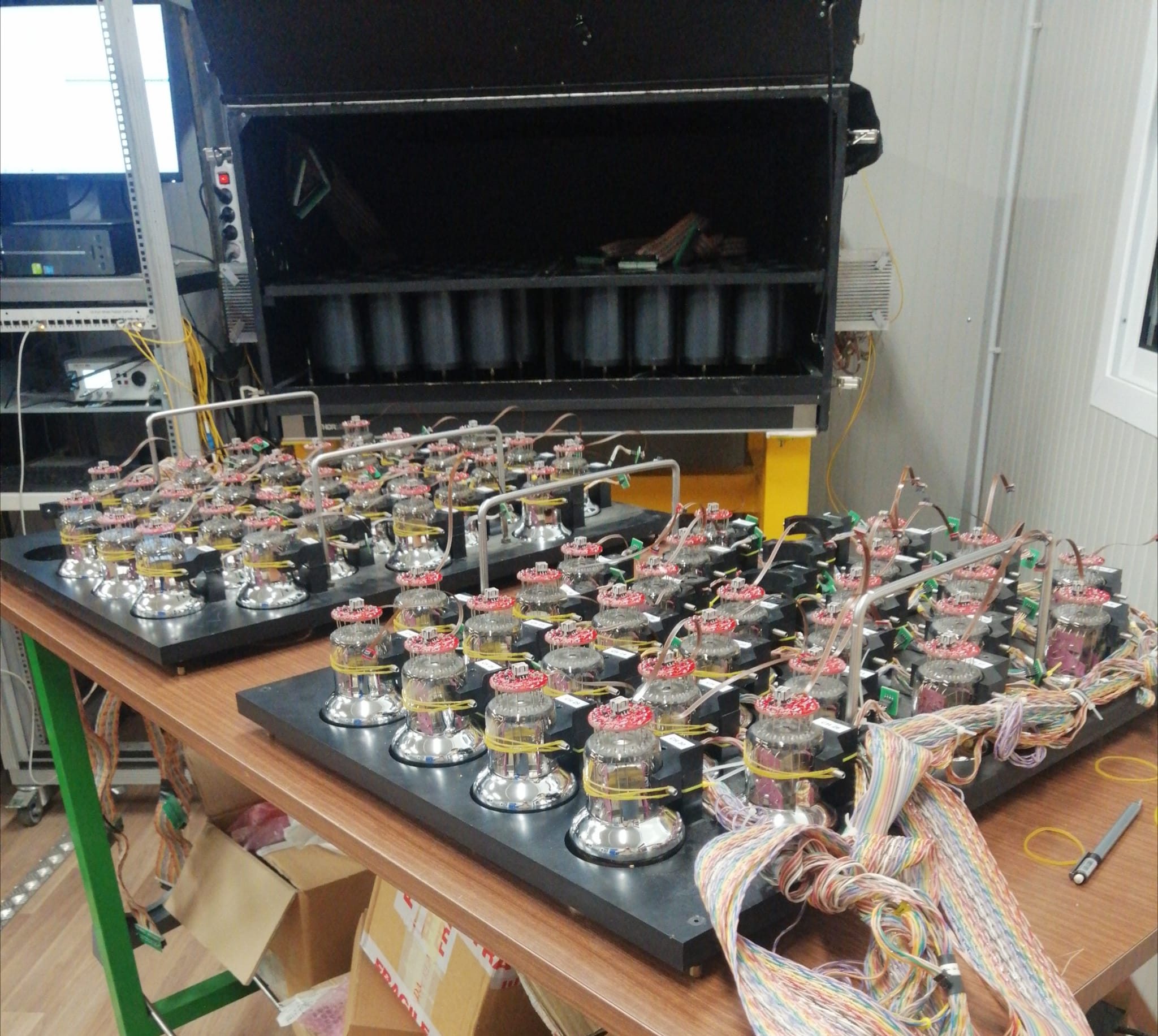}
\caption{Dark-Box apparatus and two trays of 31 R14374-02 PMTs corresponding to two DOMs. The details can be found in~\cite{Bozza:2016dtf}.}
\label{fig:scheme}
\end{figure}
To perform the measurements, a specific tool called Dark-Box \cite{Bozza:2016dtf} was used. The Dark-Box is a wooden box equipped with removable trays designed to hold the PMTs being tested. An electrical cabling system is used to connect the PMTs to the data acquisition system located outside the box, while maintaining the same time signal delay for all PMTs. The data acquisition system is designed to replicate the electronics system of the DOMs and includes a control board equipped with two signal collection boards to interface with each set of 31 PMTs. The control boards are synchronised at the subnanosecond level through the White Rabbit protocol \cite{moreira2011precise}. To illuminate all PMTs at single photon electron level, a picosecond pulsed laser and a calibrated optical splitting system are used. The Dark-Box can accommodate up to 62 PMTs per measurement session and generates for each PMT a report with the number of dark noise counts, high voltage tuning for a specific gain, and response properties of spurious pulses such as prepulses, afterpulses, and delayed pulses. To simulate a downsized detection unit made of two DOMs in a laboratory environment, the Dark-Box employs a picosecond laser with tunable repetition rate as source. The output of the laser is split by a 1-to-62 optical fiber splitter that guarantees an intensity splitting accuracy better than one percent. A typical measurement takes about 12 hours to be completed.
The apparatus had been previously used to conduct a large-scale set of 5000 measurements on the R12199-02 model \cite{oldpmts}. The results obtained with the Dark-Box are aggregated and summarised statistically.
 The PMTs are tested in the same state as when they would be integrated into the DOMs, equipped with their bases and coated with the same insulating varnish used for the bases. More information on the mechanics, electronics, laser calibration system, and the performance of the DarkBox can be found in reference \cite{Bozza:2016dtf}.
\subsection{Gain and High Voltage calibration}
\label{gain}
The nominal high voltage (HV) required to achieve a gain of $3\times10^6$, defined as the ratio between anode and photocatode currents, is provided by the manufacturer for each PMT. However, in KM3NeT, PMTs are used in pulse mode, where most of the detected pulses are due to single photoelectrons. 
A tunable threshold discriminator on the digital base used on each PMT of KM3NeT provides a rectangular signal whose duration is called time over threshold (ToT). This value is sent to the Central Logic Board (CLB) of the DOM for digitisation.
\begin{figure}[htbp]
\centering
\includegraphics[width=.45\textwidth]{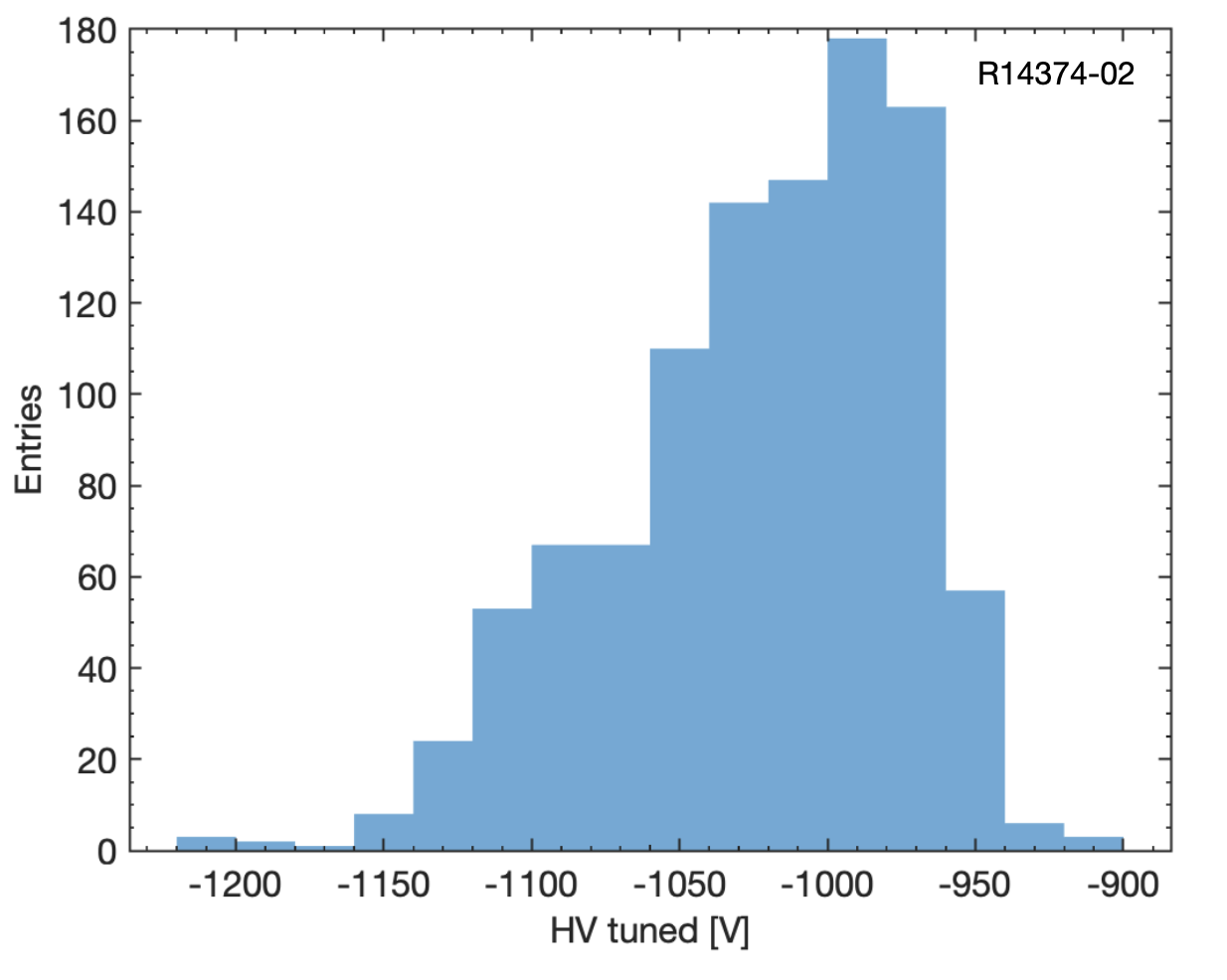}
\qquad
\includegraphics[width=.45\textwidth]{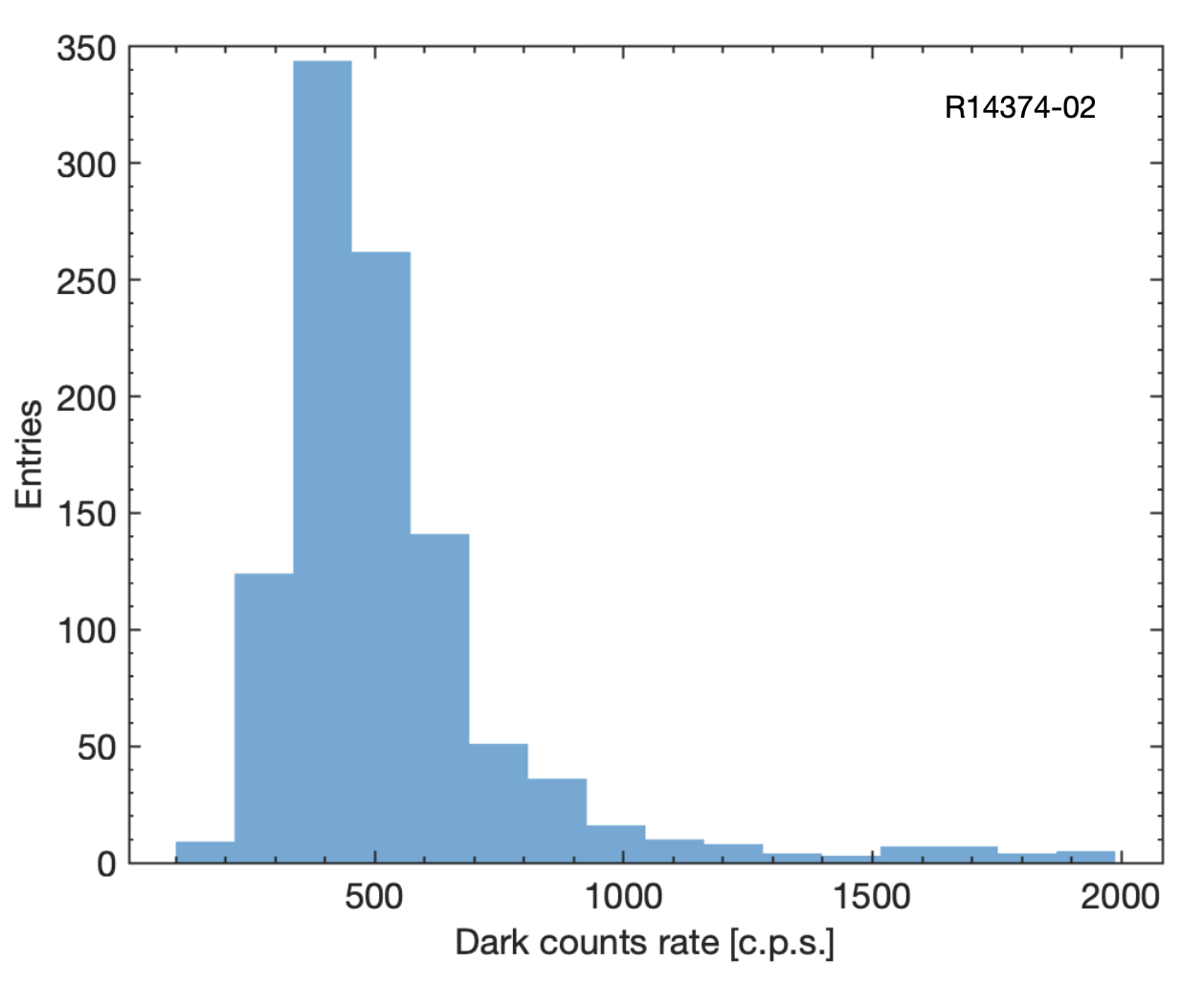}
\caption{Left: high voltage tuning results. Right: the distribution of the dark count rates (cps) is shown. \label{HVandDR}}
\label{HVplots}
\end{figure}
The method for determining the HV tuning involves adjusting the high voltage to ensure that the photomultiplier tubes exhibit the same time over threshold (ToT) peak time. This approach is identical to the one used in the study on the R12199-02 \cite{oldpmts}. The peak value of the ToT distribution at 26.4 ns for a spe signal at 0.3 spe threshold was used to set the HV value for each PMT, and these values are referred to as the "tuned HV."  In Figure \ref{HVandDR} (left) the HV values obtained after the tuning process in the Dark-Box are shown.

\subsection{Dark count rates}
Dark counts can significantly affect the performance of a PMT by generating a background signal that can mask the signals of interest. Therefore, it is important to characterise the dark count rate of PMTs to ensure their proper operation in a detector. The dark count rate is affected by several factors, including the photocathode area, the photocathode material, and the natural radioactivity in the structure of the PMT. Bialkali photo cathodes, which are used in KM3NeT PMTs, have a low dark count rate per unit area (100 cps/cm²) wich is common for high-quality, low-noise PMTs . The natural radioactivity in the structure of the PMT can also contribute to the dark count rate, and the most important components are usually $^{40}$K and Th contained in the glass envelope. The measured distribution for 1000 PMTs is shown in Figure \ref{HVandDR} (right), the mean value obtained is 530 counts per second at the tuned HV value.
The result was obtained under ambient conditions representative of typical human comfort, with diurnal temperature fluctuations ranging approximately from \(20\,^\circ\mathrm{C}\) to \(25\,^\circ\mathrm{C}\). Despite these variations, the outcome remains well within the experimental specifications.

\subsection{Measurement of PMT time characteristics and of spurious pulses}
Spurious pulses that are correlated in time with expected PMT responses include prepulses, delayed pulses, and afterpulses. 
The presence of spurious pulses can significantly degrade the time resolution of the PMT causing errors in time measurements and leading to false triggers. Therefore, measuring the probability of occurrence of spurious pulses is an important step in characterising PMT performance. 
To measure the fraction of spurious pulses and estimate the PMT time performance, the PMTs were illuminated with a 470 nm laser operating in the single photoelectron regime, with a frequency of 20 kHz. Data from runs of 10 minutes were collected and analysed. 

\subsubsection{Afterpulses} 
Afterpulses are additional noise pulses that occur after the main response of a photomultiplier tube to a detected light event. They can be classified as either early or late afterpulses. Early afterpulses are caused by light emitted from the various stages of the dynodes structure, which can reach the photocatode and generate more photoelectrons. Typically, PMTs exhibit early afterpulses within 10-80 nanoseconds after the primary pulse. Late afterpulses are caused by residual gases within the PMTs that are ionised by the passage of electrons through the space between the photocatode and the first dynode, as well as through the multiplier structure. The positive ions that are generated can drift backward and some can find their way back to the photocatode. The time taken for the ions to return to the photocatode can range from hundreds of nanoseconds to tens of microseconds and depends on various factors, such as the type of ions, their location, and the supply voltage.
\begin{figure}
\begin{center}
\includegraphics[scale=0.40]{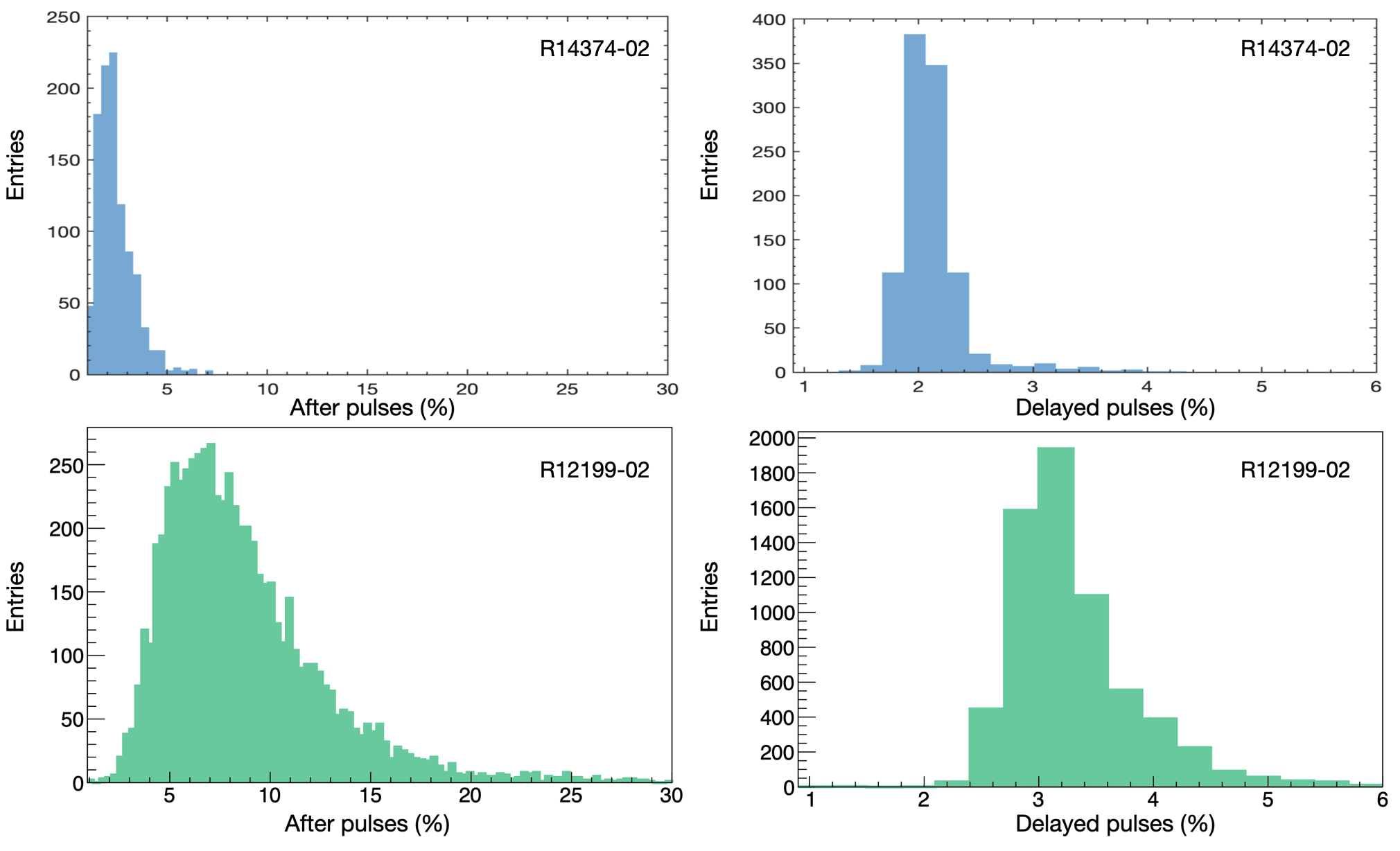}
\end{center}
\caption{\label{comparison1}Measurements of the rate of delayed pulses (right) and after pulses (left) for a set of 1000 PMTs R14374-02 (top row) and, for comparison, for the
R12199-02 model as reported in \cite{oldpmts} (bottom row).}
\end{figure}
In a 3-inch PMT, early afterpulses occur within approximately 20 nanoseconds after the first hit. However, due to the limitations of the KM3NeT front-end electronics, direct measurement of early afterpulses is not possible when consecutive hits have a time difference of less than 26 nanoseconds. 
The first photon hits are defined as pulses detected in a window of 200~ns around the expected arrival time of the PMT signal.
The percentage of afterpulses can be calculated by dividing the number of first hits in the $[T_{\text{peak}} + 100.5 \text{ ns}, T_{\text{peak}} + 10  \mu\text{s}]$ time window by the total number of first hits, where $T_{\text{peak}}$ corresponds to the centre of the maximum bin (transit time peak). The histogram of the afterpulse fraction measured for the entire 1000 PMT sample is shown in Figure~\ref{comparison1} (left) where the distribution at the top refers to the R14374-02 and shows the improvement with respect to the R12199-02 distribution at the bottom.
\subsubsection{Delayed pulses}
Delayed pulses are due to the elastic scattering of photoelectrons on the first dynode and are detected at a time that is about twice the transit time from the photocathode to the first dynode, causing a bump at around 35 ns in the falling edge of the main peak. The percentage of delayed pulses can be calculated by dividing the number of first hits in the $[T_{\text{peak}} + 15.5 \ \text{ ns}, T_{\text{peak}} + 60.5 \ \text{ ns}]$ time window by the total number of first hits. The measured distribution of delayed pulses for the R14374-02 model (Figure \ref{comparison1} (top) ) shows lower values and is narrower the distribution for the R12199-02 model (Figure \ref{comparison1} (bottom)).
\begin{figure}
\begin{center}
\includegraphics[scale=0.40]{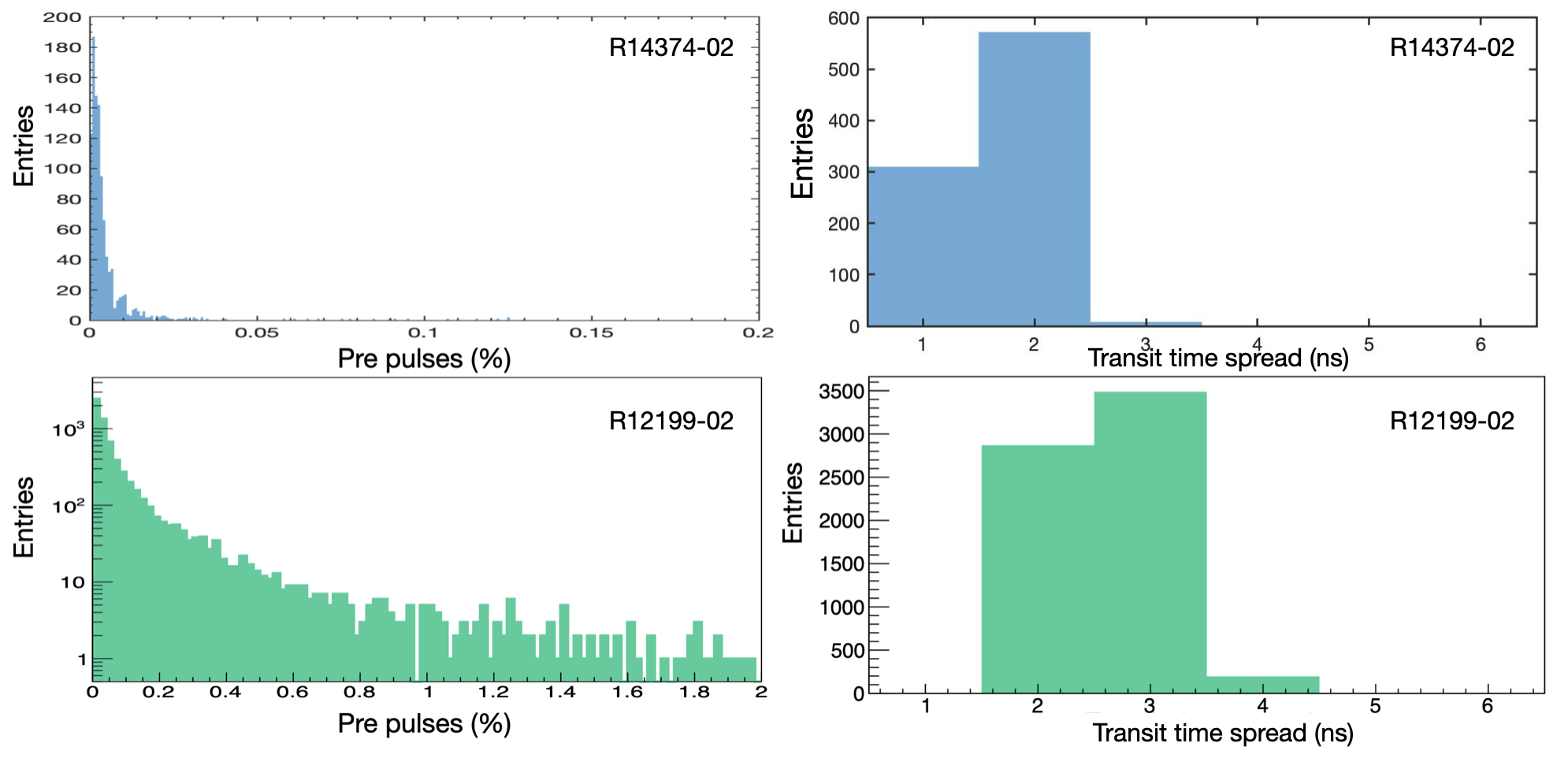}
\end{center}
\caption{\label{comparison2}Measurements of the rate of prepulses and transit time spread for a set of 1000 PMTs R14374-02 (top row) and, for comparison, for the
R12199-02 model as reported in \cite{oldpmts} (bottom row).}
\end{figure}

\subsubsection{Prepulses} 
Prepulses in photomultiplier tubes are spurious pulses that occur before the main PMT pulse. These prepulses can be a problem in applications where low background signals need to be measured accurately. The main cause of prepulses is the emission of a photoelectron at the first dynode or in structures after the photocatode. Non-uniform field emission can also lead to prepulses. In a PMT, the electric field is used to accelerate the photoelectrons from the cathode to the first dynode. If the electric field is not uniform across the surface of the cathode, some photoelectrons may be accelerated more than others, leading to a distribution of arrival times at the first dynode. Prepulses can be reduced by using high-quality cathode materials, minimising contamination of the cathode surface, and ensuring that the electric field is uniform across the cathode surface. Additionally, some PMT manufacturers have developed special coatings or treatments for the cathode to reduce prepulses.
The percentage of prepulses for this study is defined as the ratio of hits within the window $[T_{\text{peak}} - 60.5 \ \text{ ns}, T_{\text{peak}} - 10.5 \ \text{ ns}]$ to the total number of first hits.
The distribution of prepulses is shown in Figure \ref{comparison2} (left), for the R14374-02 (top) and for the R12199-02 (bottom). Note the change of scale between bottom and top plot: prepulses are significantly reduced in the R14374-02.
\subsubsection{Photoelectron transit time}
Time characteristics of PMTs are measured by
detecting and analysing the so called first photon hits, i.e. pulses detected in the window [T0,
T0 + 200 ns], where T0 is the calculated arrival time of laser photons on the PMT surface. First hits must have no hits before them in the defined time window.
The earliest high peak of the distribution corresponds to the PMT transit time, having already subtracted the travel
time of photons in the optical path and the electronic time latency between the laser trigger signal
and the laser optical output. These optical and electronics delays are fixed and measured during the
DarkBox setting-up \cite{Bozza:2016dtf}.
The value of the transit time is determined as the centre of the bin with the maximum counts.
The transit time spread is defined as the FWHM of the main peak. It is determined by counting
the number of bins for which the content value is greater than half of the maximum bin content. The final distribution of TTS over the PMT set is shown in Figure \ref{comparison2}.

\section{Conclusions}
The apparatus described in reference \cite{migliozzi2023scanning} was used to investigate the quantum efficiency of a group of 200 photomultiplier tubes of type Hamamatsu R14374-02. The result indicates that the quantum efficiency curve of the new model is consistent with the one measured in \cite{oldpmts}. A radial dishomogeneity through a linear photocatode scan partially documented in the literature \cite{hamahandbook,fotonis}, was observed. To investigate the radial behaviour over a group of ten R12199-02 and ten R14374-02 PMTs, the circular symmetry of the effect was analysed. The results show better homogeneity in the radial dependence of the quantum efficiency for the new model, as well as a slightly higher QE value in the centre. Also the time properties have been measured. A substantial and evident difference was observed between the two models. The distributions of both afterpulses and delayed pulses favour the new model because the dispersion of the distributions is narrower. The same apparatus previously employed in reference \cite{oldpmts} was used to measure the time properties of the new R14374-02 model on a set of 1000 PMTs, and the results showed a noticeable improvement in the time characteristics of this model compared to the previous one. The performances described in the manuscript meet the requirements of the experiment and suggests that the new model of 3-inch PMT currently in use in the current KM3NeT detectors will improve the overall performance of the entire detector.

\section{Acknowledgements}
The authors acknowledge the financial support of:
KM3NeT-INFRADEV2 project, funded by the European Union Horizon Europe Research and Innovation Programme under grant agreement No 101079679;
Funds for Scientific Research (FRS-FNRS), Francqui foundation, BAEF foundation.
Czech Science Foundation (GAƒaR 24-12702S);
Agence Nationale de la Recherche (contract ANR-15-CE31-0020), Centre National de la Recherche Scientifique (CNRS), Commission Europ\'eenne (FEDER fund and Marie Curie Program), LabEx UnivEarthS (ANR-10-LABX-0023 and ANR-18-IDEX-0001), Paris \^Ile-de-France Region, Normandy Region (Alpha, Blue-waves and Neptune), France,
The Provence-Alpes-Cote d'Azur Delegation for Research and Innovation (DRARI), the Provence-Alpes-Cote d'Azur region, the Bouches-du-Rhone Departmental Council, the Metropolis of Aix-Marseille Provence and the City of Marseille through the CPER 2021-2027 NEUMED project,
The CNRS Institut National de Physique Nucleaire et de Physique des Particules (IN2P3);
Shota Rustaveli National Science Foundation of Georgia (SRNSFG, FR-22-13708), Georgia;
This work is part of the MuSES project which has received funding from the European Research Council (ERC) under the European Union‚ Horizon 2020 Research and Innovation Programme (grant agreement No 101142396).
This work was supported by the European Research Council, ERC Starting grant \emph{MessMapp}, under contract no. $949555$.
The General Secretariat of Research and Innovation (GSRI), Greece;
Istituto Nazionale di Fisica Nucleare (INFN) and Ministero dell' Universit{\`a} e della Ricerca (MUR), through PRIN 2022 program (Grant PANTHEON 2022E2J4RK, Next Generation EU) and PON R\&I program (Avviso n. 424 del 28 febbraio 2018, Progetto PACK-PIR01 00021), Italy; IDMAR project Po-Fesr Sicilian Region az. 1.5.1; A. De Benedittis, W. Idrissi Ibnsalih, M. Bendahman, A. Nayerhoda, G. Papalashvili, I. C. Rea, A. Simonelli have been supported by the Italian Ministero dell'Universit{\`a} e della Ricerca (MUR), Progetto CIR01 00021 (Avviso n. 2595 del 24 dicembre 2019); KM3NeT4RR MUR Project National Recovery and Resilience Plan (NRRP), Mission 4 Component 2 Investment 3.1, Funded by the European Union ‚Äì NextGenerationEU,CUP I57G21000040001, Concession Decree MUR No. n. Prot. 123 del 21/06/2022;
Ministry of Higher Education, Scientific Research and Innovation, Morocco, and the Arab Fund for Economic and Social Development, Kuwait;
Nederlandse organisatie voor Wetenschappelijk Onderzoek (NWO), the Netherlands;
The grant ‚AstroCeNT: Particle Astrophysics Science and Technology Centre, carried out within the International Research Agendas programme of the Foundation for Polish Science financed by the European Union under the European Regional Development Fund; The program: Excellence initiative-research university for the AGH University in Krakow; The ARTIQ project: UMO-2021/01/2/ST6/00004 and ARTIQ/0004/2021;
Ministry of Research, Innovation and Digitalisation, Romania;
Slovak Research and Development Agency under Contract No. APVV-22-0413; Ministry of Education, Research, Development and Youth of the Slovak Republic;
MCIN for PID2021-124591NB-C41, -C42, -C43 and PDC2023-145913-I00 funded by MCIN/AEI/10.13039/501100011033 and by ‚ÄúERDF A way of making Europe‚Äù, for ASFAE/2022/014 and ASFAE/2022 /023 with funding from the EU NextGenerationEU (PRTR-C17.I01) and Generalitat Valenciana, for Grant AST22\_6.2 with funding from Consejer\'{\i}a de Universidad, Investigaci\'on e Innovaci\'on and Gobierno de Espa\~na and European Union - NextGenerationEU, for CSIC-INFRA23013 and for CNS2023-144099, Generalitat Valenciana for CIDEGENT/2018/034, /2019/043, /2020/049, /2021/23, for CIDEIG/2023/20, for CIPROM/2023/51 and for GRISOLIAP/2021/192 and EU for MSC/101025085, Spain;
Khalifa University internal grants (ESIG-2023-008, RIG-2023-070 and RIG-2024-047), United Arab Emirates;
The European Union's Horizon 2020 Research and Innovation Programme (ChETEC-INFRA - Project no. 101008324).

\end{document}